\documentclass[%
 reprint,
showpacs,
 amsmath,amssymb,
 aps,
pra,
]{revtex4-1}

\usepackage{amsmath}
\usepackage{amssymb}
\usepackage{epsfig}
\usepackage{graphicx}
\usepackage{dcolumn}
\usepackage{bm}
\usepackage[pdftex,bookmarksnumbered,bookmarksopen,colorlinks,citecolor=blue,linkcolor=blue]{hyperref}

\begin{document}

\preprint{APS/123-QED}

\title{Universal methods for extending any entanglement witness from the bipartite to
the multipartite case}
\author{Bang-Hai Wang$^{1,2}$}%
\author{Hai-Ru Xu$^{1,2}$}
\author{Simone Severini$^2$}%

\affiliation{%
$^1$School of Computers, Guangdong University of Technology, Guangzhou 510006, People's Republic of China\\
$^2$Department of Computer Science, and Department of Physics and Astronomy, University College London, Gower Street, WC1E 6BT London, United Kingdom
}%

\date{ \today}

\begin{abstract}
Any bipartite entanglement witness $W$ can be written as $W=c_{\sigma}I-\sigma$, where $\sigma$ is a quantum state, $I$ is the
identity matrix, and $c_{\sigma}$ is a non-negative number. We present a general method to extend the given entanglement witness to multipartite cases via purification, partial purification, and direct tensor of the quantum state $\sigma$. Our methods extend $\sigma$ but leave the parameter $c_{\sigma}$ untouched.  This is very valuable since the parameter is generally not easy to compute.
\end{abstract}
\maketitle
\section{Introduction}

It is well established that entanglement plays an essential role in many
applications of quantum information science \cite{Horodecki09,Guhne09}. The
detection of entanglement has become one of the central problems in the field. The
notion of an entanglement witness (in short, witness), which is formulated in terms of a
positive, but not completely positive map via the Jamio{\l }kowski-Choi
isomorphism \cite{Jam}, is arguably the most powerful method for entanglement
detection. An observable $W=W^{\dag}$ is said to be a \emph{witness} if:
\emph{(i)} it has non-negative expectation value for an arbitrary separable
state; and \emph{(ii)} it has at least one negative eigenvalue (see,
\emph{e.g.}, \cite{Brandao05}). A witness is said to be \emph{weakly optimal
}if its expectation value vanishes on at least one product state
\cite{Badziag13,Wang13}.

There is substantial literature on the topic. Bipartite witnesses,
\emph{i.e.}, witnesses for bipartite quantum systems, have been exhaustively
studied in a number of works
\cite{Brandao05,Terhal00,Lewenstein00,Gurvits02,Doherty04,Sperling09,Hou10a,Chruscinski09,Chruscinski10}. However, multipartite witnesses for partial and genuine entanglement are
more difficult to approach
\cite{Loock03,Bourennane04,Eisert07,Chruscinski08,Sperling13}. Of course, such
witnesses are important because multipartite entanglement has been shown to be
an essential resource in a variety of contexts, including applications in
quantum computing \cite{Vidal03}, interferometry \cite{Leibfried04}, and in
metrological tasks \cite{Gross10}.  We refer the reader to Ref. \cite{Chruscinski14} for a recent and extensive review on entanglement witnesses.

In the present paper, we consider the general form of a witness, $W=c_{\sigma
}I-\sigma$, where $c_{\sigma}$ is a non-negative real number, $I$ is the
identity matrix, and $\sigma$ is a quantum state. Given $W$, we propose methods for purifying, partially purifying, or directly
tensor the quantum state $\sigma$, and in this way extending the witness from
the bipartite to the multipartite case. Our methods modify $\sigma$ but leave
the parameter $c_{\sigma}$ unchanged. This is valuable since $c_{\sigma}$ is
in general not easy to compute.

The remainder of the paper is organized as follows. In Section II, we introduce
a general form of bipartite entanglement witnesses from density matrices of states. In Section III, we extend
any witness in bipartite to witnesses in multipartite by purification and by
partial purification. In Section IV, we show the extension by extending
states to mixed states. Section V is a summary.


\section{Preliminaries}

For our purposes, we can consider a finite dimensional composite Hilbert space
$\mathcal{H}=\mathcal{H}_{1}\otimes\mathcal{H}_{2}\otimes\cdots\otimes
\mathcal{H}_{n}$. Let $\sigma$ be a density matrix for such a system. The
quantum state $\sigma$ is said to be \emph{separable} if it can be written as
\begin{equation}
\sigma=\sum_{k}p_{k}|\psi_{1}^{k}\rangle\langle\psi_{1}^{k}|\otimes|\psi
_{2}^{k}\rangle\langle\psi_{2}^{k}|\otimes\cdots\otimes|\psi_{n}^{k}%
\rangle\langle\psi_{n}^{k}|, \label{Separable-form}%
\end{equation}
where $p_{k}$ is a probability distribution and each $|\psi_{i}^{k}\rangle$ is
a pure state of $\mathcal{H}_{i}$, for $i=1,2,...,n$. If a quantum state
$\rho$ cannot be written as the form of Eq. (\ref{Separable-form}), it is
referred to as \emph{multipartite entangled}.

On the basis of this definition, a \emph{multipartite entanglement witness},
$W\in\mathcal{H}$, is a Hermitian operator such that: \emph{(i)}
$\text{tr}(W\sigma)\geq0$ for all separable states $\sigma$; \emph{(ii)}
$\text{tr}(W\rho)<0$ for at least one state $\rho$.

Wang and Long \cite{Wang11,Wang13} showed that any (possibly unnormalized)
bipartite witness $W\in\mathcal{H}_{A}\otimes\mathcal{H}_{B}$ can be written
as
\begin{equation}
W=\rho-c_{\rho}I, \label{ew-form0}%
\end{equation}
where $\rho$ is a (separable) density matrix and
\begin{equation}
\lambda_{0\rho}<c_{\rho}\leq c_{\rho}^{max} \label{cInterval1}%
\end{equation}
is a real number related to $\rho$. Here, $\lambda_{0\rho}$ is the smallest
eigenvalue of $\rho$. The parameter $c_{\rho}^{max}$ is the maximum $c_{\rho}$
for which $W$ is a witness, and it is given by
\begin{equation}
c_{\rho}^{max}=\inf_{\parallel|\mu_{A}\rangle\parallel=1,\parallel|\mu
_{B}\rangle\parallel=1}\langle\mu_{A}\mu_{B}|\rho|\mu_{A}\mu_{B}\rangle,
\end{equation}
where $|\mu_{A}\mu_{B}\rangle$ is any unit product state. A witness
$W=\rho-c_{\rho}I$ is \emph{weakly optimal} if and only if $c_{\rho}=c_{\rho
}^{max}$.


\textbf{Remark 1.} Let $W=\rho-c_{\rho}I$ be a witness such that $\rho$ has spectral
decomposition $\rho=\sum_{i=0}^{k}\lambda_{i\rho}|e_{i}\rangle\langle e_{i}|$,
with orthonormal basis $\{|e_{1}\rangle,|e_{2}\rangle,...,|e_{k}\rangle\}$. We
have one of the following two cases: \emph{(i)} ($\rho$ has zero nullity) The
basis spans $\mathcal{H}_{AB}$ and $|e_{0}\rangle$ is an entangled state
corresponding to the minimum eigenvalue $\lambda_{0\rho}\neq0$; \emph{(ii)}
($\rho$ has nonzero nullity) The complementary space of the basis is a
\emph{completely entangled subspace} (for short, \emph{CES}). A CES does not
contain any product state. \cite{Wallach02,Parthasarathy04,Cubitt08,Walgate08}.







Similarly to Refs. \cite{Wang11,Wang13}, we can obtain the \emph{dual
witnesses' form} of Eq. (\ref{ew-form0}),
\begin{equation}
W=c_{\rho}I-\rho\label{ew-form00},%
\end{equation}
where $\rho$ is a (separable) density matrix and
\begin{equation}
c_{\rho}^{min}\leq c_{\rho}<\lambda_{M\rho} \label{cInterval2}%
\end{equation}
is a real number related to $\rho$. The maximum eigenvalue of $\rho$ is
denoted by $\lambda_{M\rho}$ is and
\begin{equation}
c_{\rho}^{min}=\sup_{\parallel|\mu_{A}\rangle\parallel=1,\parallel|\mu
_{B}\rangle\parallel=1}\langle\mu_{A}\mu_{B}|\rho|\mu_{A}\mu_{B}%
\rangle\label{ew-form1}%
\end{equation}
is the minimum $c_{\rho}$ such that $W$ is a witness; $|\mu_{A}\mu_{B}\rangle$
is any unit product state.

\textbf{Remark 2.} If $W=c_{\rho}I-\rho$ is a witness with spectral decomposition of $\rho
=\sum_{i=0}^{k}\lambda_{i\rho}|e_{i}\rangle\langle e_{i}|$, the eigenspace of
the maximum eigenvalue $\lambda_{M\rho}$ is a CES.

It is easy to conclude that any multipartite witness can be constructed from a
(separable) state of the form given in Eq. (\ref{ew-form0}) or Eq.
(\ref{ew-form00}).


\section{Purification and partial purification}

\subsection{Purification}

Purification is a fundamental tool \cite{Nielsen00}. We recall it for the sake
of completeness. Let $\rho_{A}$ be a state in Hilbert space $\mathcal{H}%
_{A}$. We introduce another system, with space denoted by $\mathcal{H}_{B}$,
and define a pure state $\psi_{AB}$ for the joint system $\mathcal{H}%
_{A}\otimes\mathcal{H}_{B}$ such that $\rho_{A}=\text{tr}_{B}(|\psi
_{AB}\rangle\langle\psi_{AB}|)$. More precisely, suppose $\rho_{A}$ has an
orthonormal decomposition $\rho_{A}=\sum_{i}p_{i}|i^{A}\rangle\langle i^{A}|$.
To \emph{purify} $\rho_{A}$, we introduce a system $\mathcal{H}_{B}$ which has
\emph{the same state space} as $\rho_{A}$ and an orthonormal basis
$|i^{B}\rangle$. We define a pure state for the combined system by $|\psi
_{AB}\rangle=\sum_{i}\sqrt{p_{i}}|i^{A}\rangle|i^{B}\rangle$. This state is
said to be a \emph{purification} of $|\psi_{AB}\rangle$. The reduced density
matrix for $\mathcal{H}_{A}$ corresponding to the state $\psi_{AB}$ is
\begin{eqnarray}
\text{tr}_B(|\psi_{AB}\rangle\langle\psi_{AB}|)  &=&\sum_{ij}\sqrt{p_{i}p_{j}%
}|i^{A}\rangle\langle j^{A}|\text{tr}(|i^{B}\rangle\langle j^{B})\\
&=&\sum_{ij}\sqrt{p_{i}p_{j}}|i^{A}\rangle\langle j^{A}|\delta_{ij}\\
&=&\sum_{i}p_{i}|i^{A}\rangle\langle i^{A}|\\
&=&\rho^{A}.
\end{eqnarray}

We are now ready to state our first result.

\textbf{Theorem 1.} If $W_{12}=c_{\sigma_{12}}I-\sigma_{12}$ is a bipartite witness,
then
\begin{equation}
W_{123}=c_{\sigma_{12}}I_{123}-|\psi_{123}\rangle\langle\psi_{123}|
\label{Multi-ew-form}%
\end{equation}
is a tripartite witness, where $|\psi_{123}\rangle$ is any purification of
$\sigma_{12}$.

In fact, this extension is the inverse process of the \emph{cascaded
structure} in \cite{Sperling13}. We give two proofs of the statement in our
language:\ The first proof is lengthy; the second one is much shorter. We
believe that both proofs are instructive.

\textbf{Lemma 1.} \cite{Nielsen00} Suppose $|AB_1\rangle$ and $|AB_2\rangle$ are two purifications of the state $\rho^A$ to a composite system $\mathcal{H}_A\otimes \mathcal{H}_B$. There exists a unitary transformation $U_B$ acting on system $\mathcal{H}_B$ such that $|AB_1\rangle=(I_A\otimes U_B)|AB_2\rangle$.


\textbf{Proof:} Let us consider the spectral decomposition $\sigma_{12}=\sum_{i}p_{i}|\psi_{12}^{i}\rangle\langle\psi_{12}^{i}|$. The state $|\psi_{123}^{\prime
}\rangle=\sum_{i}\sqrt{p_{i}}|\psi_{12}^{i}\rangle|e_{3}^{i}\rangle$ is a
purification of $\sigma_{12}$, where $|e_{3}^{i}\rangle$ is any orthonormal
basis states in $\mathcal{H}_{3}$. Suppose
\begin{eqnarray}
&&c_{|\psi_{123}^{\prime}\rangle\langle\psi_{123}^{\prime}|}^{min}\nonumber\\
&=&\max_{\parallel|\mu_{1}^{\prime}\rangle\parallel=1,\parallel|\mu_{2}^{\prime
}\rangle\parallel=1,\parallel|\mu_{3}^{\prime}\rangle\parallel=1}\langle
\mu_{1}^{\prime}\mu_{2}^{\prime}\mu_{3}^{\prime}|\psi_{123}^{\prime}%
\rangle\langle\psi_{123}^{\prime}|\mu_{1}^{\prime}\mu_{2}^{\prime}\mu
_{3}^{\prime}\rangle\nonumber\\
\\
&=&\langle\mu_{1}\mu_{2}\mu_{3}|\psi_{123}^{\prime}\rangle\langle\psi
_{123}^{\prime}|\mu_{1}\mu_{2}\mu_{3}\rangle\\
&=&rr^{\ast}%
\end{eqnarray}
and $|\mu_{3}\rangle=\sum_{i}t_{i}|e_{3}^{i}\rangle$, where $r$ and $t_{i}$
are complex numbers. Then,%
\begin{eqnarray}
r  &=&\langle\mu_{1}\mu_{2}|\langle\mu_{3}|\sum_{i}\sqrt{p_{i}}|\psi_{12}%
^{i}\rangle|e_{3}^{i}\rangle\\
&=&\langle\mu_{1}\mu_{2}|\sum_{j}t_{j}^{\ast}\langle e_{3}^{j}|\sum_{i}%
\sqrt{p_{i}}|\psi_{12}^{i}\rangle|e_{3}^{i}\rangle\\
&=&\langle\mu_{1}\mu_{2}|\sum_{i}\sqrt{p_{i}}|\psi_{12}^{i}\rangle
t_{i}^{\ast}.
\end{eqnarray}
It follows that
\[
\sum_{i}\frac{\sqrt{p_{i}}\langle\mu_{1}\mu_{2}|\psi_{12}^{i}\rangle}{r}%
t_{i}^{\ast}=1.
\]
Let $t_{i}=\frac{\sqrt{p_{i}}\langle\mu_{1}\mu_{2}|\psi_{12}^{i}\rangle}{r}$.
Since $\sum_{i}t_{i}t_{i}^{\ast}=1$, we have
\[
|\mu_{3}\rangle=\sum_{i}\frac{\sqrt{p_{i}}}{r}\langle\mu_{1}\mu_{2}|\psi
_{12}^{i}\rangle|e_{3}^{i}\rangle.
\]
Similarly, suppose
\[
c_{\sigma_{12}}^{min}=\langle\mu_{1}\mu_{2}|\sigma_{12}|\mu_{1}\mu_{2}%
\rangle.
\]
We can write%
\begin{eqnarray}
&&c_{\sigma_{12}}^{min}\nonumber\\
&=&\langle\mu_{1}\mu_{2}|(\sum_{i}p_{i}|\psi_{12}%
^{i}\rangle\langle\psi_{12}^{i}|)|\mu_{1}\mu_{2}\rangle\\
&=&\sum_{i}\sqrt{p_{i}}\langle\mu_{1}\mu_{2}|\psi_{12}^{i}\rangle\sum
_{j}\sqrt{p_{j}}\langle\psi_{12}^{j}|\mu_{1}\mu_{2}\rangle\langle e_{3}%
^{i}|e_{3}^{j}\rangle\\
&=&r\sum_{i}\sqrt{p_{i}}\langle\psi_{12}^{i}|\mu_{1}\mu_{2}\rangle(\langle
e_{3}^{i}|\underbrace{\sum_{j}\frac{\sqrt{p_{j}}}{r}\langle\mu_{1}\mu_{2}%
|\psi_{12}^{j}\rangle|e_{3}^{j}\rangle}_{=|\mu_{3}\rangle})\nonumber\\
\\
&=&r\sum_{i}\sqrt{p_{i}}\langle\psi_{12}^{i}|\mu_{1}\mu_{2}\rangle(\langle
e_{3}^{i}|\mu_{3}\rangle)\\
&=&r\langle\psi_{123}^{\prime}|\mu_{1}\mu_{2}\mu_{3}\rangle\\
&=&\langle\mu_{1}\mu_{2}\mu_{3}|\psi_{123}^{\prime}\rangle\langle\psi
_{123}^{\prime}|\mu_{1}\mu_{2}\mu_{3}\rangle\\
&=&c_{|\psi_{123}^{\prime}\rangle\langle\psi_{123}^{\prime}|}^{min},
\end{eqnarray}
since
\begin{eqnarray}
&&c_{|\psi_{123}^{\prime}\rangle\langle\psi_{123}^{\prime}|}^{min}\nonumber\\
&=&\max_{\parallel|\mu_{1}^{\prime}\rangle\parallel=1,\parallel|\mu_{2}^{\prime
}\rangle\parallel=1,\parallel|\mu_{3}^{\prime}\rangle\parallel=1}|\langle
\mu_{1}^{\prime}\mu_{2}^{\prime}\mu_{3}^{\prime}|\psi_{123}^{\prime}%
\rangle|^{2}\\
&=&\max_{\parallel|\mu_{3}^{\prime}\rangle\parallel=1}\left(  \max
_{\parallel|\mu_{1}^{\prime}\rangle\parallel=1,\parallel|\mu_{2}^{\prime
}\rangle\parallel=1}|\langle\mu_{3}^{\prime}|\psi_{123}^{\prime|\mu
_{1}^{\prime}\mu_{2}^{\prime}\rangle}\rangle|^{2}\right)  ,
\end{eqnarray}
where
\[
|\psi_{123}^{\prime|\mu_{1}^{\prime}\mu_{2}^{\prime}\rangle}\rangle
:=\langle\mu_{1}^{\prime}\mu_{2}^{\prime}|\psi_{123}^{\prime}\rangle.
\]
Since $|\psi_{123}^{\prime}\rangle\langle\psi_{123}^{\prime}|$ is a pure
state, $\lambda_{M|\psi_{123}^{\prime}\rangle\langle\psi_{123}^{\prime}|}=1$
and $\lambda_{M\sigma_{12}}\leq\lambda_{M|\psi_{123}^{\prime}\rangle
\langle\psi_{123}^{\prime}|}$. Since $W_{12}=c_{\sigma_{12}}I-\sigma_{12}$ is
a bipartite witness, $c_{\sigma_{12}}^{min}\leq c_{\sigma_{12}}<\lambda
_{M\sigma_{12}}$. Thus,
\[
c_{|\psi_{123}^{\prime}\rangle\langle\psi_{123}^{\prime}|}^{min}%
=c_{\sigma_{12}}^{min}\leq c_{\sigma_{12}}<\lambda_{M\sigma_{12}}\leq
\lambda_{M|\psi_{123}^{\prime}\rangle\langle\psi_{123}^{\prime}|}.
\]
By Eq. (\ref{ew-form00}), Eqs. (\ref{cInterval2}) and Eq. (\ref{ew-form1}),
\[
W_{123}=c_{\sigma_{12}}I_{123}-|\psi_{123}^{\prime}\rangle\langle\psi
_{123}^{\prime}|
\]
is a witness. By Lemma 1, for any purification $|\psi_{123}\rangle$ of
$\sigma_{12}$, Eq. (\ref{Multi-ew-form}) holds.
\hfill$\blacksquare$

\bigskip

\textbf{Proof:}
The proof uses two points:\

\emph{(i)} Suppose $\mu_{i}$ is any unit pure state of a system with Hilbert
space $\mathcal{H}_{i}$, for $i=1,2,3$. We have
\begin{eqnarray}
&&\langle\mu_{1}\mu_{2}\mu_{3}|W_{123}|\mu_{1}\mu_{2}\mu_{3}\rangle \nonumber\\
&=&c_{\sigma_{12}}-tr_{12}(tr_{3}(|\psi_{123}\rangle\langle\psi_{123}||\mu
_{3}\rangle\langle\mu_{3}|)|\mu_{1}\mu_{2}\rangle\langle\mu_{1}\mu_{2}|)\nonumber\\
\\
&\geq& c_{\sigma_{12}}-tr_{12}(tr_{3}(|\psi_{123}\rangle\langle\psi
_{123}|)|\mu_{1}\mu_{2}\rangle\langle\mu_{1}\mu_{2}|)\\
&=&c_{\sigma_{12}}-tr_{12}(\sigma_{12}|\mu_{1}\mu_{2}\rangle\langle\mu_{1}%
\mu_{2}|)\\
&=&tr(W_{12}|\mu_{1}\mu_{2}\rangle\langle\mu_{1}\mu_{2}|)\\
&\geq&0,
\end{eqnarray}
since $W_{12}$ is a bipartite witness.

\emph{(ii)}$\ $By Eq. (\ref{cInterval2}), $W_{123}<0$ since $c_{\sigma_{12}%
}<1$.

Combining together \emph{(i)} and \emph{(ii)}, $W_{123}$ is a tripartite witness.
\hfill$\blacksquare$

\textbf{Corollary 1.}
If $W_{12}=c_{\sigma_{12}}I-\sigma_{12}$ is a bipartite witness, then
\begin{equation}
W_{12\cdots n}=c_{\sigma_{12}}I_{12\cdots n}-|\psi_{123}\rangle\langle
\psi_{123}|\otimes|\psi_{4}\rangle\langle\psi_{4}|\otimes\cdots\otimes
|\psi_{n}\rangle\langle\psi_{n}| \label{Multi-ew-gen1}
\end{equation}
is a $n$-partite witness, where $|\psi_{123}\rangle$ is any purification of
$\sigma_{12}$ and $|\psi_{i}\rangle\langle\psi_{i}|$ is a pure state in
$\mathcal{H}_{i}$.

We illustrate our results for the case of the isotropic qubit state. It is
known that
\[
\sigma_{q}=\left(
\begin{array}
[c]{cccc}%
\frac{1+q}{4} & 0 & 0 & \frac{q}{2}\\
0 & \frac{1-q}{4} & 0 & 0\\
0 & 0 & \frac{1-q}{4} & 0\\
\frac{q}{2} & 0 & 0 & \frac{1+q}{4}%
\end{array}
\right)
\]
is the (separable) density matrix of
\[
\sigma_{q}=q|\psi\rangle\langle\psi|+(1-q)I/4,
\]
where $|\psi\rangle=\frac{1}{\sqrt{2}}(|00\rangle+|11\rangle)$ and
$0<q<\frac{1}{3}$.

By computing $c_{\sigma_{q}}^{min}=\frac{1+q}{4}$ \cite{Wang11},
\begin{equation}
W_{12}=\frac{1+q}{4}I-\sigma_{q} \label{Weaklyoptimal}%
\end{equation}
is a bipartite witness for $0<q<\frac{1}{3}$. It works for
\[
\pi_{p}=p|\psi\rangle\langle\psi|+(1-p)I/4,
\]
where $p>\frac{1}{3}$. Since the maximum eigenvalue of $\sigma_{q}$ is
$\lambda_{M\sigma}=\frac{1+3q}{4}$,
\begin{equation}
W_{12}=c_{\sigma_{q}}I-\sigma_{q} \label{Not-weakly-optimal}%
\end{equation}
is a bipartite witness for $\frac{1+q}{4}\leq c_{\sigma_{q}}<\frac{1+3q}{4}$.

The spectral decomposition for $\sigma_{q}$ is
\begin{eqnarray}
\sigma_{q}&=&\frac{1-q}{4}|\psi_{12}^{0}\rangle\langle\psi_{12}^{0}|+\frac
{1-q}{4}|\psi_{12}^{1}\rangle\langle\psi_{12}^{1}|\nonumber\\
&&+\frac{1-q}{4}|\psi_{12}%
^{2}\rangle\langle\psi_{12}^{2}|+\frac{1+3q}{4}|\psi_{12}^{3}\rangle
\langle\psi_{12}^{3}|,
\end{eqnarray}
where $|\psi_{12}^{0}\rangle=|10\rangle$ and $|\psi_{12}^{1}\rangle
=|01\rangle$ are separable, while $|\psi_{12}^{2}\rangle=\frac{1}{\sqrt{2}%
}(|00\rangle-|11\rangle)$ and $|\psi_{12}^{3}\rangle=\frac{1}{\sqrt{2}%
}(|00\rangle+|11\rangle)$ are entangled.

A purification of $\sigma_{q}$ in $C^{2}\otimes C^{2}\otimes C^{4}$ is
\begin{eqnarray}
|\psi_{123}\rangle&=&\sqrt{\frac{1-q}{4}}|\psi_{12}^{0}\rangle|0\rangle
+\sqrt{\frac{1-q}{4}}|\psi_{12}^{1}\rangle|1\rangle\nonumber\\
&&+\sqrt{\frac{1-q}{4}}%
|\psi_{12}^{2}\rangle|2\rangle+\sqrt{\frac{1+3q}{4}}|\psi_{12}^{3}%
\rangle|3\rangle,
\end{eqnarray}
where $\{|i\rangle\}_{i=0}^{3}$ is the orthogonal basis in $C^{4}$. Hence,
\[
W_{123}=\frac{1+q}{4}I_{123}-|\psi_{123}\rangle\langle\psi_{123}|
\]
is a tripartite witness in $C^{2}\otimes C^{2}\otimes C^{4}$ for $0<q<\frac
{1}{3}$.




\subsection{Partial purification}

The dimension of $\mathcal{H}_{3}$ is the same as the dimension of
$\mathcal{H}_{1}\otimes\mathcal{H}_{2}$ (or it is equal to the rank of
$\sigma_{12}$) because of the demand of purification. There exists a large
gap. There also exists a restriction to the dimension of $\mathcal{H}_{3}$ for
the extension. One would hope that $\mathcal{H}_{3}$ can be of any dimension.
Here we need to extend the notion of purification. The joint system $(A,B)$
will be on a space $\mathcal{H}_{A}\otimes\mathcal{H}_{B}$.

Suppose the spectral decomposition for $\rho_{A}$ is
\[
\rho_{A}=\lambda_{0}|\phi_{0}^{A}\rangle\langle\phi_{0}^{A}|+\lambda_{1}%
|\phi_{1}^{A}\rangle\langle\phi_{1}^{A}|+\cdots+\lambda_{M}|\phi_{M}%
^{A}\rangle\langle\phi_{M}^{A}|,
\]
with $\lambda_{0}\leq\lambda_{1}\leq\cdots\leq\lambda_{M}$, where $\lambda
_{0}$ and $\lambda_{M}$ are the minimum and maximum eigenvalue, respectively.
To purify $\rho_{A}$ to the system $B$, whose dimension is less than the rank
of $\rho_{A}$, we define a (unnormalized) pure state $|\phi_{AB}\rangle$ for
the joint system such that
\[
|\phi_{AB}\rangle=\sum_{i}\sqrt{\lambda_{i}}|\phi_{i}^{A}\rangle|e_{i}%
^{B}\rangle,
\]
This state is said to be a \emph{partial purification} of $\rho^{A}$. The
states $|e_{i}^{B}\rangle$ form an orthonormal basis in $\mathcal{H}_{B}$.
Moreover, $\lambda_{i}$ and $|\phi_{i}^{A}\rangle$ are selected from the
eigenvalues and eigenvectors of $\rho_{A}$. When $\lambda_{M}$ and $|\phi
_{M}^{A}\rangle$ (respectively, $\lambda_{0}$ and $|\phi_{0}^{A}\rangle$) are
selected, the state $|\phi_{AB}\rangle$ is said to be a \emph{partial
purification} with maximum eigenvalue (respectively minimum eigenvalue).

Generally, a partial purification $|\phi_{AB}\rangle$ is unnormalized and
$\text{tr}_{B}(|\phi_{AB}\rangle\langle\phi_{AB}|)\leq\rho^{A}$. For example,
if the spectral decomposition for $\sigma_{12}\in\mathbb{C}^{2}\otimes
\mathbb{C}^{2}$ is
\[
\sigma_{12}=p_{0}|\phi_{12}^{0}\rangle\langle\phi_{12}^{0}|+p_{1}|\phi
_{12}^{1}\rangle\langle\phi_{12}^{1}|+p_{2}|\phi_{12}^{2}\rangle\langle
\phi_{12}^{2}|+p_{3}|\phi_{12}^{3}\rangle\langle\phi_{12}^{3}|,
\]
where $|\phi_{12}^{i}\rangle$ is the orthonormal basis state in $\mathbb{C}%
^{2}\otimes\mathbb{C}^{2}$. Partial purifications of $\sigma_{12}$ are
\[
|\phi_{123}\rangle=\sqrt{p_{0}}|\phi_{12}^{0}\rangle|0\rangle+\sqrt{p_{1}%
}|\phi_{12}^{1}\rangle|1\rangle,
\]
and
\[
|\phi_{123}^{\prime}\rangle=\sqrt{p_{0}}|\phi_{12}^{0}\rangle|0\rangle
+\sqrt{p_{3}}|\phi_{12}^{3}\rangle|1\rangle.
\]

\textbf{Theorem 2.} If $W_{12}=c_{\sigma_{12}}I-\sigma_{12}$ is a bipartite witness,
\begin{equation}
W_{123}=c_{\sigma_{12}}I_{123}-|\phi_{123}\rangle\langle\phi_{123}|
\label{Multi-ew-form1}%
\end{equation}
is a tripartite witness, where the (unnormalized) pure state $|\phi
_{123}\rangle$ is a partial purification of $\sigma_{12}$ with maximum eigenvalue.

\textbf{Proof:}
We proceed as follows:\

\emph{(i)} By Theorem 1,
\[%
\begin{tabular}
[c]{lll}%
$c_{|\phi_{123}\rangle\langle\phi_{123}|}^{min}\leq c_{\sigma_{12}}^{min}$ &
and & $c_{|\phi_{123}\rangle\langle\phi_{123}|}^{min}\leq c_{\sigma_{12}}.$%
\end{tabular}
\]

\emph{(ii)} By the definition of partial purification, the maximum eigenvalue
of $|\phi_{123}\rangle$ is equal to the sum of eigenvalues selected to
partially purify. Since we select the maximum eigenvalue $\lambda
_{M\sigma_{12}}$, we have then
\[
\lambda_{M\sigma_{12}}\leq\lambda_{M|\phi_{123}\rangle\langle\phi_{123}|}.
\]
Since $W_{12}=c_{\sigma_{12}}I-\sigma_{12}$ is a bipartite witness,
\[%
\begin{tabular}
[c]{lll}%
$c_{\sigma_{12}}<\lambda_{M\sigma_{12}}$ & and & $c_{\sigma_{12}}%
<\lambda_{M|\phi_{123}\rangle\langle\phi_{123}|}.$%
\end{tabular}
\]

By \emph{(i)} and \emph{(ii)},
\[
c_{|\phi_{123}\rangle\langle\phi_{123}|}^{min}\leq c_{\sigma_{12}}%
<\lambda_{M|\phi_{123}\rangle\langle\phi_{123}|}.
\]

Combining together Eq. (\ref{ew-form00}), Eq. (\ref{cInterval2}) and Eq.
(\ref{ew-form1}), we can observe that Eq. (\ref{Multi-ew-form1}) is a
tripartite witness.
\hfill$\blacksquare$

 \textbf{Corollary 2.}
If $W_{12}=c_{\sigma_{12}}I-\sigma_{12}$ is a bipartite witness then
\begin{equation}
W_{123}=c_{\sigma_{12}}^{\prime}I_{123}-|\phi_{123}\rangle\langle\phi_{123}|%
\end{equation}
is a tripartite witness, where $c_{\sigma_{12}}\leq c_{\sigma_{12}}^{\prime
}<\lambda_{M|\phi_{123}\rangle\langle\phi_{123}|}$ and the (unnormalized) pure
state $|\phi_{123}\rangle$ is a partial purification of $\sigma_{12}$ with
maximum eigenvalue.

As for Theorem 1, we can also give a simpler proof of Theorem
2.

\textbf{Proof:}
As for Theorem 1, the proof uses two points:\

\emph{(i)} Suppose $|\mu_{i}\rangle$ is any unit pure state of the system
$\mathcal{H}_{i}$ for $i=1,2,3$. We have
\begin{align}
&  \langle\mu_{1}\mu_{2}\mu_{3}|W_{123}|\mu_{1}\mu_{2}\mu_{3}\rangle
\nonumber\\
&  =c_{\sigma_{12}}-tr_{12}(tr_{3}(|\psi_{123}\rangle\langle\psi_{123}%
||\mu_{3}\rangle\langle\mu_{3}|)|\mu_{1}\mu_{2}\rangle\langle\mu_{1}\mu
_{2}|)\nonumber\\
&  \geq c_{\sigma_{12}}-tr_{12}(tr_{3}(|\psi_{123}\rangle\langle\psi
_{123}|)|\mu_{1}\mu_{2}\rangle\langle\mu_{1}\mu_{2}|)\\
&  \geq c_{\sigma_{12}}-tr_{12}(\sigma_{12}|\mu_{1}\mu_{2}\rangle\langle
\mu_{1}\mu_{2}|)\\
&  =tr(W_{12}|\mu_{1}\mu_{2}\rangle\langle\mu_{1}\mu_{2}|)\\
&  \geq0,
\end{align}
since $W_{12}$ is a bipartite witness.

\emph{(ii)} We have $W_{123}<0$ since, by Eq. (\ref{cInterval2}),
$c_{\sigma_{12}}<1$.

By \emph{(i)} and \emph{(ii)} together, $W_{123}$ is a tripartite witness.
\hfill$\blacksquare$

 \textbf{Corollary 3.}
If $W_{12}=c_{\sigma_{12}}I_{12}-\sigma_{12}$ is a bipartite witness then
\begin{equation}
W_{12\cdots n}=c_{\sigma_{12}}I_{12\cdots n}-|\phi_{123}\rangle\langle
\phi_{123}|\otimes|\phi_{4}\rangle\langle\phi_{4}|\otimes\cdots\otimes
|\phi_{n}\rangle\langle\phi_{n}| \label{Multi-ew-gen2}%
\end{equation}
is a $n$-partite witness, where $|\phi_{123}\rangle$ is any partial
purification of $\sigma_{12}$ with maximum eigenvalue and $|\phi_{i}%
\rangle\langle\phi_{i}|$ is a pure state in $\mathcal{H}_{i}$.

Although it seems that Theorem 2 is not as powerful as Theorem 1, we can construct a series of witnesses. Suppose that rank$(\sigma
_{12})=R$ is the rank of $\sigma_{12}\in\mathcal{H}_{12}$. Let
dim$(\mathcal{H}_{3})=d_{3}<R\,\ $be the dimension of $\mathcal{H}_{3}$, which
is the space to which we want to extend. We can construct exactly
\[
\sum_{i=1}^{d_{3}}\frac{(R-1)!}{(R-i)!(i-1)!}\cdot\frac{d_{3}!}{(d_{3}-i)!}=n
\]
tripartite witnesses by partial purification because the maximum eigenvalue
and eigenvector must be selected.

To extend Eq. (\ref{Weaklyoptimal}) to tripartite witnesses in three copies of
$\mathbb{C}^{2}$ (\emph{i.e.}, $\mathbb{C}^{2}\otimes\mathbb{C}^{2}%
\otimes\mathbb{C}^{2}$), we can construct
\[
W_{123}^{1}=\frac{1+q}{4}I_{123}-|\psi_{123}\rangle\langle\psi_{123}|
\]
with one of the partial purification of $\sigma_{q}$ with maximum eigenvalue
\[
|\psi_{123}\rangle=\sqrt{\frac{1+3q}{4}}|\psi_{12}^{3}\rangle|0\rangle
+\sqrt{\frac{1-q}{4}}|\psi_{12}^{2}\rangle|1\rangle.
\]
Also,
\[
W_{123}^{2}=\frac{1+q}{4}I_{123}-|\psi_{123}^{\prime}\rangle\langle\psi
_{123}^{\prime}|
\]
with one of the partial purification of $\sigma_{q}$ with maximum eigenvalue
\[
|\psi_{123}^{\prime}\rangle=\sqrt{\frac{1+3q}{4}}|\psi_{12}^{3}\rangle
|0\rangle+\sqrt{\frac{1-q}{4}}|\psi_{12}^{1}\rangle|1\rangle.
\]
Finally,
\[
W_{123}^{3}=\frac{1+q}{4}I_{123}-|\psi_{123}^{\prime\prime}\rangle\langle
\psi_{123}^{\prime\prime}|,
\]
where%
\[
|\psi_{123}^{\prime\prime}\rangle=\sqrt{\frac{1+3q}{4}}|\psi_{12}^{3}%
\rangle|1\rangle+\sqrt{\frac{1-q}{4}}|\psi_{12}^{0}\rangle|0\rangle,
\]
and so on.

\subsection{Bipartite witnesses}

In standard quantum mechanics, there is no entanglement and no witness for a
single system. However, it is interesting that we can consider a witness
$W_{1}=0$ for a single system. We can extend $W_{1}$ to bipartite witnesses. Then,%

\[
W_{1}=0=\frac{1}{2}I-\frac{1}{2}I
\]
in a $2$-level system, can be extended to a bipartite witness $W_{12}$ in
$\mathbb{C}^{2}\otimes\mathbb{C}^{2}$. We purify
\[
\frac{1}{2}I=\frac{1}{2}(|0\rangle\langle0|+|1\rangle\langle1|)
\]
to the pure state
\[
|\psi\rangle=\frac{1}{\sqrt{2}}(|0\rangle\otimes|0\rangle+|1\rangle
\otimes|1\rangle)
\]
in the composite system%
\[
W_{12}=\frac{1}{2}I_{12}-|\psi\rangle\langle\psi|.
\]
We can also purify $\frac{1}{2}I$ to
\[
|\phi\rangle=\frac{1}{\sqrt{2}}(|0\rangle\otimes|1\rangle+|1\rangle
\otimes|0\rangle),
\]
and obtain the bipartite witness%
\[
W_{12}=\frac{1}{2}I_{12}-|\phi\rangle\langle\phi|.
\]

\section{Mixed states extension}

We begin the section with a question:\ can we extend bipartite witnesses by
transforming states into mixed states in Eq. (\ref{ew-form00})? The following
result answers the question:

\textbf{Theorem 3.}If $W_{12}=c_{\sigma_{12}}I_{12}-\sigma_{12}$ is a bipartite
witness then
\begin{eqnarray}
W_{12\cdots n}&=&c_{\sigma_{12}}I_{12\cdots n}-\sigma_{12}\otimes\frac
{1}{\lambda_{M\sigma_{3}}}\sigma_{3}\otimes\cdots\nonumber\\
&&\otimes\frac{1}%
{\lambda_{M\sigma_{i}}}\sigma_{i}\otimes\cdots\otimes\frac{1}{\lambda
_{M\sigma_{n}}}\sigma_{n}
\end{eqnarray}
is a $n$-partite witness, where $\sigma_{i}$ is any
(normalized, mixed or pure) state in $\mathcal{H}_{i}$
with maximum eigenvalue $\lambda_{M\sigma_{i}}$.

\textbf{Proof:}
\emph{(i)} Firstly,
\begin{eqnarray}
&&c_{\sigma_{12}\otimes\frac{1}{\lambda_{M\sigma_{3}}}\sigma_{3}\otimes
\cdots\otimes\frac{1}{\lambda_{M\sigma_{n}}}\sigma_{n}}^{min} \nonumber\\
&=&\max_{\parallel|\mu_{1}^{\prime}\rangle\parallel=1,\parallel|\mu_{2}^{\prime
}\rangle\parallel=1,\cdots\parallel|\mu_{n}^{\prime}\rangle\parallel=1}%
\langle\mu_{1}^{\prime}\mu_{2}^{\prime}\cdots\mu_{n}^{\prime}|\nonumber\\
&&\sigma_{12}\otimes\frac{1}{\lambda_{M\sigma_{3}}}\sigma_{3}\otimes
\cdots\otimes\frac{1}{\lambda_{M\sigma_{n}}}\sigma_{n}|\mu_{1}^{\prime}\mu
_{2}^{\prime}\cdots\mu_{n}^{\prime}\rangle\\
&=&\max_{\parallel|\mu_{1}^{\prime}\rangle\parallel=1,\parallel|\mu
_{2}^{\prime}\rangle\parallel=1}\langle\mu_{1}^{\prime}\mu_{2}^{\prime}%
|\sigma_{12}|\mu_{1}^{\prime}\mu_{2}^{\prime}\rangle\cdot\langle e_{3}%
|\frac{1}{\lambda_{M\sigma_{3}}}\sigma_{3}|e_{3}\rangle\cdot\nonumber\\
&&\cdots\cdot\langle e_{i}|\frac{1}{\lambda_{M\sigma_{i}}}\sigma_{i}%
|e_{i}\rangle\cdot\cdots\cdot\langle e_{n}|\frac{1}{\lambda_{M\sigma_{n}}%
}\sigma_{n}|e_{n}\rangle\\
&=&c_{\sigma_{12}}^{min},
\end{eqnarray}
where $|e_{i}\rangle\langle e_{i}|$ is the eigenvector corresponding to the
maximum eigenvalue of $\sigma_{i}$, with $3\leq i\leq n$.

\emph{(ii)} Then, since in these cases the maximum eigenvalue of $\frac
{1}{\lambda_{M\sigma_{i}}}\sigma_{i}$ is $1$, we have%
\[
\lambda_{M\sigma_{12}}=\lambda_{M\sigma_{12}\otimes\frac{1}{\lambda
_{M\sigma_{3}}}\sigma_{3}\otimes\cdots\otimes\frac{1}{\lambda_{M\sigma_{i}}%
}\sigma_{i}\otimes\cdots\otimes\frac{1}{\lambda_{M\sigma_{n}}}\sigma_{n}}.
\]
By \emph{(i)} and \emph{(ii)}, $W_{12\cdots n}$ is an $n$-partite witness.
\hfill$\blacksquare$

Can we extend the bipartite witness by purification or partial purification in
the form given by Eq. (\ref{ew-form0}) [the dual form of Eq.
(\ref{ew-form00})]? The answer is negative, since the minimum eigenvalue is $0$
after the process of purification and then the space spanned by the
eigenvectors generally is not a CES if we purify the state $\sigma_{12}$. The
following extension, however, can be done by extending pure states to mixed
states from both Eqs. (\ref{ew-form0}) and (\ref{ew-form00}).

\textbf{Corollary 4.} If $W_{12}=c_{\sigma_{12}}I_{12}-\sigma_{12}$ is a bipartite
witness then
\begin{equation}
W_{12\cdots n}=c_{\sigma_{12}}I_{123}-\sigma_{12}\otimes I_{3}\otimes
\cdots\otimes I_{n}\label{Multi-ew-form2}%
\end{equation}
is a $n$-partite witness in $\mathcal{H}_{1}\otimes\mathcal{H}_{2}%
\otimes\cdots\mathcal{H}_{n}$.

If $\rho_{12}$ is the entangled state witnessed by $W_{12}$, $tr(W_{12\cdots n}\rho_{12\cdots n})<0$ by Corollary 4, where $\rho_{12\cdots n}=\rho_{12}\otimes\rho_{3}\cdots\rho_{i}\cdots\otimes\rho_{n}$ and $\rho_{i}$
is any state in $\mathcal{H}_{i}$. This result indicates that all states can
be witnessed by $W_{12\cdots n}$, and that this is given by the tensor product
of any state witnesses by $W_{12}$ and any state in $\mathcal{H}_{i}$.

By Corollary 4, we can thus extend Eq. (\ref{Weaklyoptimal}) to
\[
W_{123}^{q}=\frac{1+q}{4}I_{123}-\sigma_{123}^{q}%
\]
where
\begin{eqnarray}
&&\sigma_{123}^{q}\nonumber\\
&=&(\frac{1-q}{4}\left(  |\psi_{12}^{0}\rangle\langle\psi
_{12}^{0}|+|\psi_{12}^{1}\rangle\langle\psi_{12}^{1}|+|\psi_{12}^{2}%
\rangle\langle\psi_{12}^{2}|\right)  \nonumber\\
&&+\frac{1+3q}{4}|\psi_{12}^{3}%
\rangle\langle\psi_{12}^{3}|)\otimes(|0\rangle\langle0|+|1\rangle\langle
1|+|2\rangle\langle2|+|3\rangle\langle3|)\nonumber\\
\end{eqnarray}
in $C^{2}\otimes C^{2}\otimes C^{4}$.

We can extend Eq. (\ref{Weaklyoptimal}) to
\begin{equation}
W_{123}^{q}=\frac{1+q}{4}I_{123}-\sigma_{123}^{q}%
\end{equation}
where%
\begin{eqnarray}
\sigma_{123}^{q}&=&(\frac{1-q}{4}\left(  |\psi_{12}^{0}\rangle\langle\psi
_{12}^{0}|+|\psi_{12}^{1}\rangle\langle\psi_{12}^{1}|+|\psi_{12}^{2}%
\rangle\langle\psi_{12}^{2}|\right) \nonumber\\
&&+\frac{1+3q}{4}|\psi_{12}^{3}%
\rangle\langle\psi_{12}^{3}|)\otimes(|0\rangle\langle0|+|1\rangle\langle1|)
\end{eqnarray}
in $C^{2}\otimes C^{2}\otimes C^{2}$.

Consider the same witness (unnormalized)
\[%
\begin{tabular}
[c]{lll}%
$W_{12}=\frac{1}{4}W_{12}^{|\psi\rangle}$  & as (normalized) & $W_{12}^{|\psi\rangle}%
=|\psi\rangle\langle\psi|^{\Gamma}$,
\end{tabular}
\]
where $\psi=\frac{1}{\sqrt{2}}(|00\rangle+|11\rangle)$ and $\Gamma$ refers to
partial transposition. We can write $W_{12}$ in the form given in Eq. (\ref{ew-form0})
\[
W_{12}=\frac{1}{4}|\psi\rangle\langle\psi|^{\Gamma}=\sigma_{12}-\frac{3}%
{16}I_{12},%
\]
where
\[
\sigma_{12}=\left(
\begin{array}
[c]{cccc}%
\frac{5}{16} & 0 & 0 & 0\\
0 & \frac{3}{16} & \frac{1}{8} & 0\\
0 & \frac{1}{8} & \frac{3}{16} & 0\\
0 & 0 & 0 & \frac{5}{16}%
\end{array}
\right)  .
\]
Hence, we have the spectral decomposition
\begin{eqnarray}
\sigma&=&\frac{1}{16}|\psi_{12}^{0}\rangle\langle\psi_{12}^{0}|+\frac{5}%
{16}|\psi_{12}^{1}\rangle\langle\psi_{12}^{1}|\nonumber\\
&&+\frac{5}{16}|\psi_{12}%
^{2}\rangle\langle\psi_{12}^{2}|+\frac{5}{16}|\psi_{12}^{3}\rangle\langle
\psi_{12}^{3}|
\end{eqnarray}
where $|\psi_{12}^{0}\rangle=\frac{1}{\sqrt{2}}(|01\rangle-|10\rangle)$ and
$|\psi_{12}^{1}\rangle=\frac{1}{\sqrt{2}}(|01\rangle+|10\rangle)$ are
entangled, while $|\psi_{12}^{2}\rangle=|00\rangle$ and $|\psi_{12}^{3}%
\rangle=|11\rangle$ are separable. We can extend $\sigma_{12}$ to
\begin{eqnarray}
\sigma_{123}&=&(\frac{1}{16}|\psi_{12}^{0}\rangle\langle\psi_{12}^{0}|+\frac
{5}{16}|\psi_{12}^{1}\rangle\langle\psi_{12}^{1}|+\frac{5}{16}|\psi_{12}%
^{2}\rangle\langle\psi_{12}^{2}|\nonumber\\
&&+\frac{5}{16}|\psi_{12}^{3}\rangle\langle
\psi_{12}^{3}|)\otimes(|0\rangle\langle0|+|1\rangle\langle1|+|2\rangle
\langle2|+|3\rangle\langle3|)\nonumber\\
\end{eqnarray}
which is full rank in $\mathbb{C}^{2}\otimes\mathbb{C}^{2}\otimes
\mathbb{C}^{4}$, and
\begin{equation}
W_{123}^{1}=\sigma_{123}-\frac{3}{16}I_{123}\label{Examplefullrank1}%
\end{equation}
is a tripartite witness. Similarly, we can extend $\sigma_{12}$ to
\begin{eqnarray}
\sigma_{123}^{\prime}&=&(\frac{1}{16}|\psi_{12}^{0}\rangle\langle\psi_{12}%
^{0}|+\frac{5}{16}|\psi_{12}^{1}\rangle\langle\psi_{12}^{1}|+\frac{5}{16}%
|\psi_{12}^{2}\rangle\langle\psi_{12}^{2}|\nonumber\\
&+&\frac{5}{16}|\psi_{12}^{3}%
\rangle\langle\psi_{12}^{3}|)\otimes(|0\rangle\langle0|+|1\rangle\langle1|),
\end{eqnarray}
which is full rank in $\mathbb{C}^{2}\otimes\mathbb{C}^{2}\otimes
\mathbb{C}^{2}$, and
\begin{equation}
W_{123}^{2}=\sigma_{123}^{\prime}-\frac{3}{16}I_{123}\label{Examplefullrank2}%
\end{equation}
is a tripartite witness.

Similar to the proof of Theorem 1, we can show $\lambda_{0\sigma
_{123}}<\frac{3}{16}\leq c_{\sigma_{123}}^{max}$ in Eq.
(\ref{Examplefullrank1}) and $\lambda_{0\sigma_{123}^{\prime}}<\frac{3}%
{16}\leq c_{\sigma_{123}^{\prime}}^{max}$ in Eq. (\ref{Examplefullrank2}), and
that Eqs. (\ref{Examplefullrank1}) and (\ref{Examplefullrank2}) give
tripartite witnesses.

Note that we can also extend $\sigma_{12}$ to tripartite
states by selecting partial bases in $\mathcal{H}_{3}$, but cannot for the extending
from the form of Eq.
(\ref{ew-form0}). The simplest extension is just the following easy but still
useful result, which can be also directly drawn from Theorem 3.

\textbf{Corollary 5.}
If $W_{12}=c_{\sigma_{12}}I_{12}-\sigma_{12}$ is a bipartite witness then
\[
W_{12\cdots n}=c_{\sigma_{12}}I_{12\cdots n}-\sigma_{12}\otimes|\psi
_{3}\rangle\langle\psi_{3}|\otimes\cdots\otimes|\psi_{n}\rangle\langle\psi
_{n}|
\]
is an $n$-partite witness, where $|\psi_{i}\rangle\langle\psi_{i}|$ is a pure
state in $\mathcal{H}_{i}$.

\section{Summary}

Based on the general form of a witness $W_{12}=c_{\sigma_{12}}I_{12}%
-\sigma_{12}$, we extend a bipartite witness to tripartite witnesses
$W_{123}=c_{\sigma_{12}}I_{123}-|\psi_{123}\rangle\langle\psi_{123}|$ by
purifying or partially purifying $\sigma_{12}$ to $|\psi_{123}\rangle
\langle\psi_{123}|$. We extend a bipartite witness $W_{12}=c_{\sigma_{12}%
}I_{12}-\sigma_{12}$ to tripartite witnesses by extending $\sigma_{12}$ to
mixed product states in $\mathcal{H}_{1}\otimes\mathcal{H}_{2}\otimes
\mathcal{H}_{3}$. For all methods, we do not need to change the parameter
$c_{\sigma_{12}}$. Our methods are universal and generalizable to extend a
bipartite witness to the multipartite case.

\bigskip

\begin{acknowledgments}
We are grateful to Fernando Brand\~{a}o and Marco
Piani for very helpful discussion and suggestions.  We thank the referee for valuable suggestions to improve the original manuscript. We carried on this work
while Wang was an academic visitor and Xu an undergraduate visiting student of
the Department of Computer Science and the Department of Physics \& Astronomy
at University College London. Wang and Xu would like to thank this institution
for the kind hospitality. This work is supported by the National Natural
Science Foundation of China under Grant No. 61272013 and the National Natural
Science Foundation of Guangdong province of China under Grant No. s2012040007302.
\end{acknowledgments}

\end{document}